\documentclass [10pt, letterpaper]{article}
\usepackage{graphics}

\newtheorem{definition}{Definition}
\newenvironment{dfn}{\begin{definition}\hspace{-.5em}{\bf
:}\rm}{\end{definition}}

\newtheorem{example}{Example}
\newenvironment{xmpl}{\begin{example}\hspace{-.5em}{\bf
:}\rm}{\end{example}}

\begin{document}

\title{
Automatic Classification of Text Databases Through Query Probing}

\author{
    Panagiotis G. Ipeirotis \\
    {\tt pirot@cs.columbia.edu}  \\
    Computer Science Dept. \\ Columbia University \\
    \and
    Luis Gravano \\
    {\tt gravano@cs.columbia.edu}  \\
    Computer Science Dept. \\ Columbia University \\
    \and
    Mehran Sahami \\
    {\tt sahami@epiphany.com}\\
    E.piphany, Inc. \\
}
\date{}

\maketitle

\begin{abstract}
Many text databases on the web are ``hidden'' behind search
interfaces, and their documents are only accessible through querying.
Search engines typically ignore the contents of such search-only
databases. Recently, Yahoo-like directories have started to manually
organize these databases into categories that users can browse to find
these valuable resources. We propose a novel strategy to automate the
classification of search-only text databases.  Our technique starts by
training a rule-based document classifier, and then uses the
classifier's rules to generate probing queries. The queries are sent
to the text databases, which are then classified based on the number
of matches that they produce for each query. We report some initial
exploratory experiments that show that our approach is promising to
automatically characterize the contents of text databases accessible
on the web.
\end{abstract}

\section{Introduction}
\label{sec:intro}

Text databases abound on the Internet. Sometimes users can browse through
their documents by following hyperlinks. In many other cases, text
databases are ``hidden'' behind search interfaces, and their documents are
only available through querying. For those databases, web search engines
cannot crawl inside, and they just index the ``front pages'', ignoring the
contents of possibly rich sources of information. One example of such a
search-only text database is the archive of a newspaper. Many newspapers do
not offer a browsing interface for past issues, but they do offer search
capabilities to retrieve old articles. This is the case, for example, for
the New York Times newspaper.

One way of facilitating the access to such searchable databases is to build
metasearchers. A metasearcher sends user queries to many search engines,
retrieves and merges the results and then returns the combined results back to
the user (e.g., \cite{article/tods/Gloss99,conf/vldb/Yu98,/conf/sigir/Callan98,
/conf/www/SelbergEtzioni95,/conf/sigmod/starts97,/conf/www/inquirus98}).
Alternatively, users can browse Yahoo-like directories to locate databases of
interest and then submit queries to these databases. Some sites have started in
the last two years to provide such services. For example,
{\em InvisibleWeb}\footnote{\texttt{http://www.invisibleweb.com/}} and
{\em SearchEngine Guide}\footnote{\texttt{http://www.searchengineguide.com/}}
classify various search engines into a hierarchical classification scheme. The
archive  of the {\em New York Times} is classified as: News $\rightarrow$
Regional $\rightarrow$ National (USA) $\rightarrow$ New York $\rightarrow$ News.
A user can then locate relevant text databases and submit queries only to them
to obtain more accurate and focused results than when searching a more general
text database. Other services (e.g.,
{\em Copernic}\footnote{\texttt{http://www.copernic.com/}}) combine the
metasearching approach with ``browsing''. Users can select a specific category
(e.g., Recipes, Newspapers, etc.) and the metasearcher then sends the user
queries to the searchable databases previously classified in the given category.

Unfortunately, existing approaches for text database classification involve
manual intervention of a human expert and do not scale. In this paper we will
describe a way of automating this classification process by issuing {\em query
probes} to the text databases. More specifically, in Section~\ref{sec:clas-def}
we define what it means to classify a text database. Then, in
Section~\ref{sec:calculate} we focus on the design of our query probing
classification strategy. Finally, in Section~\ref{sec:evaluation} we present
some initial experiments over web databases.

\paragraph{Related Work} Query probing has been used recently for characterizing
different properties of text databases. Manually constructed query probes have
been used in~\cite{article/jucs/ProFusion96} for the classification of text
databases. \cite{conf/sigmod/CallanCD99} probe text databases with queries to
determine an approximation of their vocabulary and associated statistics. This
technique requires retrieving the documents in the query results for further
analysis. Finally, guided query probing has been used
in~\cite{conf/coopis/Meng99} to determine sources of heterogeneity in the
algorithms used to index and search locally at each text database.

\section{Classification of Text Databases}
\label{sec:clas-def}

In this section we will describe two basic approaches for classifying text
databases. One approach classifies a database into one category when the
database {\em contains a substantial number of documents} in this category.
The other approach classifies a database into one category when {\em the
majority of its documents} are in this category.

\begin{xmpl}
\label{ex:intuition}
Consider two databases $D_1,~D_2$ with 1,000 and 10,000,000 documents,
respectively, and a topic category {\em ``Health.''} Suppose that $D_1$ contains
900 documents about health while $D_2$ contains 200,000 such documents.
Our decision whether to classify $D_1$ and $D_2$ in the {\em ``Health''}
category will ultimately depend on how users will take advantage of our
classification and the databases. Some users might prefer a
``focus-oriented'' classification (i.e., might be looking for text databases
having mostly documents about health and little else). Such users might not want
to process documents outside of their topic of interest, and might then prefer
that database $D_1$ be classified in the ``Health'' category (90\% of its
documents are on health). In contrast, $D_2$ should not be classified in that
category. Although $D_2$ has a large number of document on health, these
documents represent only a small fraction of the database (i.e., 2\%). Hence, it
is likely that our ``focus-oriented'' users would be exposed to non-health
documents while exploring $D_2$. Alternatively, other users might be looking for
text databases having a sufficiently large number of documents on health. It
might be unimportant for such users what else is at each database. These users
might then prefer $D_2$ to be classified in the {\em ``Health''} category
because of its large number of documents on health (i.e., 200,000). $D_1$ (with
900 documents on health) might or might not be classified in that category,
depending on what we consider a ``sufficient large'' number of documents.
\end{xmpl}

Consider a set of categories $C_1,\ldots,C_k$ and a text database $D$ that
we want to classify in one or more of these categories. Each of $D$'s documents
has been classified in one of the categories $C_1,\ldots,C_k$ that we use to
classify $D$. Given this classification of the documents in $D$ we can
compute a vector $C=\left(n_1,\ldots,n_k\right)$, which indicates the
number of documents $n_i$ in category $C_i$, for $i=1,\ldots,k$. Vector $C$
is a good summary of the contents of database $D$ and we will use it to classify
the database, as we describe next. As illustrated in
Example~\ref{ex:intuition} above, to categorize databases we need to capture how
``focused''
$D$ is and how many documents it contains for a given category.
For this we define the following two metrics.

\begin{dfn}
Consider a text database $D$ and a category $C_i$. Then the {\em coverage of $D$
for $C_i$} is the number of documents in $D$ in category $C_i$, $n_i$.
The {\em specificity of $D$ on $C_i$} is the fraction of documents in $D$ in
category $C_i$:

\begin{eqnarray*}
\mbox{\em Coverage}(D,C_i) &=& n_i \\
\mbox{\em Specificity}(D,C_i) &=& \frac{n_i}{|D|}
\end{eqnarray*}
\end{dfn}
\emph{Specificity} defines how ``focused'' a database is on a given category.
One problem with the definition above is that we do not always know the number
of documents in a database. We will discuss how we can approximate this value in
Section~\ref{sec:calculate}. \emph{Coverage} defines the ``absolute'' amount of
information that a database contains about a specific category. An alternative
definition for coverage could divide $n_i$ by the total number of documents in
all databases. This would capture what fraction of the existing documents in
category $C_i$ are present in a given database. Although this definition is
interesting, it has the undesirable property of depending on a universe of known
databases. On the Internet, databases come and go constantly so this definition
would make the resulting classification scheme that we describe quite unstable.
Moreover, since the \emph{Coverage} value would have the same normalizing
constant for all databases, excluding this factor will have no bearing on the
relative ranking of databases by their coverage of a certain topic.

Using the definitions above, each database $D$ has a specificity and a coverage
value for each category. We can use these values to decide how to classify $D$
into one or more of the categories. As described above, we could classify a
database into one category when the majority of the documents it contains are of
a specific category. Our classification could alternatively be based on the
number of documents of a specific category that a database contains.

\begin{dfn} \label{dfn:clasif-def} Consider a database $D$ and a category
$C_i$ and let $\tau_s,\tau_c\geq 0$ to be two pre-specified thresholds. Then
$D$ is in category $C_i$ according to a {\em ``coverage-oriented''
classification} if {\em Coverage$(D,C_i) \geq \tau_c$}. Similarly, $D$ is
in category $C_i$ according to a {\em ``specificity-oriented''
classification} if {\em Specificity$(D,C_i) \geq \tau_s$}.
\end{dfn}

\addtocounter{example}{-1}

\begin{xmpl}{\bf ~(cont.)} Consider the two databases $D_1,~D_2$ described
above, and the category {\em ``Health.''} Using Definition~\ref{dfn:clasif-def},
{\em Coverage$(D_1,$``Health''$) = 900$}, since $D_1$ has 900 documents on
health. Similarly, {\em Coverage$(D_2,$``Health''$) = 200,000$}. If threshold
$\tau_c$ for our {\em ``coverage-oriented'' classification} is set to, say,
10,000, then $D_2$ will be classified in category ``Health'' while $D_1$
will not, since it does not have a sufficient large number of documents in
this category. Analogously, {\em Specificity$(D_1,$``Health''$) =
\frac{900}{1000}=0.9$} while {\em Specificity$(D_2,$``Health''$) =
\frac{200,000}{10,000,000}=0.02$}. If threshold $\tau_s$ for our
{\em ``specificity-oriented'' classification} is set to, say, 0.3 then $D_1$
will be classified in category ``Health'' while $D_2$ will not, since it is not
sufficiently focused on health and holds too many documents in other
categories.
\end{xmpl}

The two alternative database classification schemes above assume that we somehow
know the number of documents that each database has in each category, which is
clearly unrealistic in most Internet settings. In effect, as discussed in the
Introduction, many times we do not have access to a database's contents other
than through a query interface. In the next section we introduce techniques for
approximating the classification of text databases in this limited-access
scenario.

\section{Classifying Databases through Probing}
\label{sec:calculate}

The previous section described how to classify a database given the number of
documents it contains in each of our categories. Unfortunately, text databases
do not export such metadata. In this section we introduce a technique to
classify text databases in the absence of any information about their contents.
Our technique starts by training a rule-based {\em document} classifier over our
categories (Section~\ref{sec:classifier}) and then uses the classifier's rules
to design a set of probing queries (Section~\ref{sec:probing}). The database
will be classified based on the number of matches returned for each of these
queries, without accessing the documents per se (Section~\ref{sec:matches}).

\subsection{Training a Document Classifier}
\label{sec:classifier}

Our technique for classifying databases over a set of categories
$C_1,\ldots,C_k$ starts by training a rule-based {\em document} classifier over
those categories. We use RIPPER, an off-the-shelf tool developed at AT\&T
Research Laboratories\cite{conf/icml/Cohen95,conf/aaai/Cohen96}. Given a set of
training, pre-classified documents, this tool returns a classifier that might
consist of rules like the following:

\begin{tabular}{ll}
\\
\texttt{Computers} & {\tt IF mac}\\
\texttt{Computers} & {\tt  IF graphics windows} \\
\texttt{Religion} & {\tt  IF god christian} \\
\texttt{Hobbies} & {\tt  IF baseball}\\
\\
\end{tabular}

The first rule indicates that if a document contains the term \texttt{mac} it
should be classified in the \emph{``Computers''} category. A document should
also be classified into that category if it has the words \texttt{graphics} and
\texttt{windows}. Similarly, if a document has the words \texttt{god} and
\texttt{christian}, it is a \emph{``Religion''} document, whereas if it has the
word \texttt{baseball}, it is a \emph{``Hobbies''} document.

Once we have trained a document classifier using a tool like RIPPER, we could
apply it to every document in a {\em database} $D$ that we want to classify.
This procedure would produce a close approximation to the
$C=\left(n_1,\ldots,n_k\right)$ vector of category frequencies for $D$
(Section~\ref{sec:clas-def}), which we could use to classify $D$ according to
Definition~\ref{dfn:clasif-def}. Unfortunately, we often do not have access to
all the documents in a database, other than indirectly through a query
interface, as discussed above. Next, we define a query probing strategy to deal
with such databases.

\subsection{Probing a Database}
\label{sec:probing}

Our goal is  to create a set of queries for each category that will retrieve
exactly the documents for that category from the database we are classifying. We
will construct these queries based on the document classifier discussed above.
To create our queries, we turn each rule into a query. The number of matches for
each query will be the number of documents in the database that satisfy the
corresponding rule. These numbers will then be used to approximate the
distribution of documents in categories within a text database, as the following
example illustrates.

\begin{xmpl} Consider a database $D$ with 500 documents, all about
\emph{``Computers,''} and suppose that our categories of interest are
\emph{``Computers,'' ``Hobbies,''} and \emph{``Religion.''} Then $D$ has
associated with it a vector $C = \left(500,0,0\right)$
(Section~\ref{sec:clas-def}), showing the distribution of documents over these
three categories. Suppose also that we have trained a rule-based document
classifier and obtained the four rules shown above for the three categories. If
we do not have access to all the documents of $D$, we can still characterize its
contents by issuing probing queries constructed from the document classifier as
discussed above. Our first probe will be the query {\tt mac}. The database
will return a result of the form ``92 documents found.'' We send a second query
{\tt graphics AND windows}. Again, we get a result like ``288 documents found.''
Queries {\tt god AND christian} and {\tt baseball} return 0 and 2 matches
respectively. From these results we conclude that $D$ has 288+92=380
``Computers'' documents, 0 ``Religion'' documents, and 2 ``Hobbies'' documents.
Thus we approximate the ideal vector $C$, with $C'=\left(380,0,2\right)$.
\end{xmpl}

RIPPER can produce either  an ordered set of rules or an unordered set of rules.
When the rules are ordered, the first rule that is satisfied by a document fires
and gives a classification for that document.  No subsequent rules are matched
against that document. We should formulate our queries properly in order to
simulate the actions of the classifier as much as possible. For example, if the
rules above were ordered rules, our second probing query would have been
{\tt graphics AND windows AND NOT mac}, to avoid retrieving any documents that
would match the first, earlier rule.

If the query interface of a database does not support the kind of queries
described above, we break these queries into smaller pieces that we can send
separately. A detailed description of this technique is beyond the scope of this
paper. For completeness, we mention that we submit the probing queries in such a
way that we can use the inclusion-exclusion principle to calculate the number of
results that would have been returned for the original queries.

A significant advantage of our probing approach is that we do not need to
retrieve documents to analyze the contents of a
database~\cite{conf/sigmod/CallanCD99}. Instead, we count only the number of
matches for these queries. Thus, in our approach we only require a database
to report the number of matches for a given query. It is common for a database
to return something like ``$X$ documents found'' before returning the actual
results.

\subsection{Using the Probing Results for Classification}
\label{sec:matches}

After the probing phase, we have calculated an approximation of the coverage of
a database for our categories. To calculate the specificity values, we would
need the size of the database $|D|$, and we approximate it by
$|D|\simeq \mathop{\sum}_{i=1}^{k}{n_i}$. This means that we will use only the
documents that are classified into the given categories to calculate the size of
the database. This approach can give poor results when there are many documents
that do not belong to any of the given categories.  In such a case, it is also
difficult to categorize this text database into the given classification scheme,
since no category will accurately reflect its contents.

An extra step that we applied to our method to improve the results is the
following. For each of the rules, we know the accuracy from the training phase
of the classifier. For example, the rule \texttt{ Computers IF mac} may have
correctly classified 90 documents and incorrectly classified 10 other documents
during the training phase, resulting in an accuracy of 0.9. We adjust our
results from the probing phase by multiplying the number of documents matched by
each rule by the accuracy of that rule. Also, for the set of rules that
classified documents into one category, we know their ``recall'', i.e., how many
documents they recalled over all the documents in this category. For example if
category Computers in the training phase had 150 documents and the rules
retrieved 100, then the recall is 0.67. This means that only this portion of all
the documents of this category were retrieved. To adjust our results further, we
divide each element of the $C'$ approximation vector with the recall for this
category. This regularization of the values $n_i$ helps account for the fact
that rules generally do not (and need not) have perfect recall on real document
databases.

\section{Initial Experiments}
\label{sec:evaluation}

Using RIPPER, we created a classifier using a collection of 20,000 newsgroup
articles from the UCI KDD archive\footnote{{\tt http://kdd.ics.uci.edu/}}. This
collection has been used in previous text categorization
experiments~\cite{tech/cmu/joachims96,book/mitchell/ml}, and is composed of
1,000 newsgroup articles from each of 20 newsgroups. We further grouped the
articles into five large groups according to their originating newsgroups:
{\em Computers} ({\tt comp.*}), {\em Science} ({\tt sci.*}), {\em Hobbies}
({\tt rec.*}), {\em Society} ({\tt alt.atheism}, {\tt talk.*}, {\tt soc.*}) and
{\em Misc} ({\tt misc.sale}). We have removed all the headers (except for the
``Subject:'' line), the e-mail addresses from the body of the articles and all
punctuation. Subsequently, we eliminated all words that appeared in fewer than 3
documents in the collection and the 100 most frequent words.
Such feature reduction is in accordance with Zipf's Law~\cite{book/zipf}, which
shows that there are many infrequently used words in document collections.  For
purposes of classification, however, such infrequent terms generally provide
little discriminating power between classes (due to their rarity), and can thus
be safely eliminated with little, if any, reduction in subsequent classification
accuracy.  Similarly, very frequent words, that often tend to appear in
virtually all articles, will also provide little ability to make classification
distinctions, and can likewise be eliminated. After this step we applied an
information theoretic feature selection
algorithm~\cite{conf/icml/Sahami96,conf/icml/Sahami97} to reduce the terms from
about 40,000 to 5,000. This algorithm eliminates features that have the least
impact on the class distribution of documents (as measured by the relative
entropy of the distribution of the document class labels conditioned on the
appearance of a given feature). Features that have little impact on the class
distribution are likely to also have little discriminating power between
classes, and can thus be eliminated without much adverse impact on the final
classification accuracy.  For training set we used a random sample of 10,000
documents and the remaining 10,000 documents were used for testing.

The initial document classifier generated by RIPPER consisted of 534 ordered
rules. Many of the rules were covering very few (one or two) examples from the
training set. These rules did not contribute much to the overall accuracy of the
document classifier, and would result in too many probing queries during the
classification stage. Thus, we decided to restrict the classifier to produce
only rules that covered at least 50 examples from the training set. This
resulted in a classifier with 29 ordered rules that included a total of 32
words. We also tried to produce a rule set that would include rules with
negations (NOT clauses).  The resulting classifier had 31 rules with much better
accuracy, but, in this case, a total of 92 words were used to form the rules.
The queries for this classifier were much longer and we opted to use the simpler
classifier (that had only 29 rules and 32 words) for the sake of query
efficiency.  The rules given in Section~\ref{sec:classifier} are, in fact,
examples of rules used by this classifier.

After constructing the classifier, we have selected four sites from
{\em InvisibleWeb}\footnote{\texttt{http://www.invisibleweb.com/}}
to test our method. These four sites are topically cohesive, and should be
classified in the same category by both the specificity- and the coverage-
oriented classification alternatives of Definition~\ref{dfn:clasif-def}:

\begin{itemize}
    \item {\em Cora\footnote{\texttt{http://www.cora.jprc.com/}}}:
    A repository of technical papers about Computer Science.
    This database should be classified under the category {\em ``Computers.''}

    \item {\em American Scientist\footnote{\texttt{http://www.amsci.org/}}}:
    An on-line version of a magazine on science and technology.
    This repository should be classified under category {\em ``Science.''}

    \item {\em AllOutdoors\footnote{\texttt{http://www.alloutdoors.com/}}}:
    A site with articles about fishing, hunting, and other outdoor activities.
    This site should be classified under category {\em ``Hobbies.''}

    \item {\em ReligionToday\footnote{\texttt{http://www.religiontoday.com/}}}:
    A site with news and discussions about religion. This site should be
    classified  under category {\em ``Society.''}

\end{itemize}

\begin{figure}[t]
\centering
\resizebox{\textwidth}{!}{
\includegraphics*[0mm,0mm][165mm,60mm]{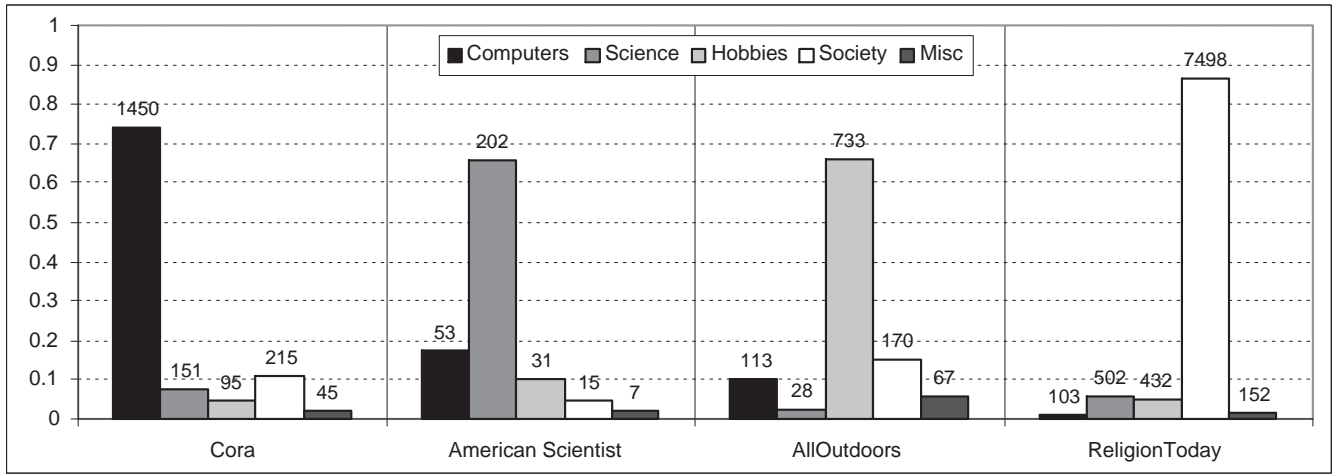}
}
\caption{Specificity and coverage values for four web-accessible databases.}
\label{fig:results}
\end{figure}

We probed these sites using the techniques described in
Section~\ref{sec:probing}. One problem that arose during the probing phase was a
limitation on the length of the queries that we could submit to the ``American
Scientist'' site. We truncated the long queries by eliminating terms that did
not cover any documents (e.g., instead of issuing a query {\tt baseball AND NOT
god}, if the query {\tt god} returned 0 results, we issue only the query {\tt
baseball}).

The results of our probing phase can be seen in Figure~\ref{fig:results}.
Consider, for example, the results for Cora. After submitting the queries for
the class Computer, the database reported 1450 matches for all the queries. For
classes Science, Hobbies, Society, and Misc, it reported 151, 95, 215, and 45
respectively. Using these coverage values we estimated specificity as in
Section~\ref{sec:clas-def}. The specificity values are depicted using the bars,
and it can be clearly seen that the results indicate that Cora is a site that is
``focused'' on Computers. Similarly for the other sites, we probed them using
the same rules. The results clearly indicate the focus of each site. For
example, if we had a threshold value for specificity of $\tau_s=0.6$, then each
site would be classified correctly. Moreover, to measure the significance of our
results, we performed a Chi-squared test comparing the distribution of the
classes for each database given by the probes to the uniform distribution.  This
test gives us a measure for how likely the skew in the class distribution
(toward the correct class) is likely to have been gotten by chance.  The
Chi-square test reveals that the skews in the class distributions for each
database are significant at the 99.9\% level. Thus, it appears that, in every
case, the probes generated by the RIPPER rules have accurately captured the
concept represented by each class of documents.

\section{Conclusions and Future Work}
\label{sec:conclusion}

In this paper, we have described a method that uses probing queries produced by
a classifier to classify a text database. We have also shown some promising
initial experiments. The method managed to identify the right category for each
database, using only the number of matches for a small set of queries and
without retrieving any documents. Our technique could also be used to
characterize web sites that offer a browsable interface as well. The only
requirement is the existence of a search interface for the local contents, which
many sites offer. By using only a small set of probe queries, we can get a
coarse idea about the contents of a web site.

Our future work includes the expansion of our strategy into a hierarchical
classification scheme. We will also explore the efficiency of our algorithm for
various indexing environments and for search interfaces that support different
sets of boolean operators. We also plan to compare our approach against an
adaptation, for the database classification problem, of the technique in
\cite{conf/sigmod/CallanCD99}. Finally, we will expand our adjustment technique
(that currently uses only the precision of each rule and the recall of each
category) to use the full set of statistics (i.e., confusion matrices) from the
document classifier. This could produce better approximations of the contents of
the search-only text databases.

{\small \bibliography{submission} }
\bibliographystyle{alpha}

\end{document}